\documentclass[preprint,prd,nofootinbib,tightenlines,groupedaddress,amsmath,amssymb]{revtex4}

\usepackage{bm}
\usepackage{color}
\usepackage{graphicx}
\allowdisplaybreaks
\usepackage{multirow}

\begin{document}


\title{CP violation in neutrino mixing  with $\delta = -\pi/2$\\ in $A_4$ Type-II seesaw model}

\author{Guan-Nan Li$^{1,3}$\footnote{lgn198741@126.com }}
\author{Xiao-Gang He$^{2,3,4}$\footnote{hexg@phys.ntu.edu.tw}}

\affiliation{${}^{1}$Department of Physics, Zhengzhou University, Zhengzhou, Henan 450001, China}
\affiliation{${}^{2}$INPAC, SKLPPC and Department of Physics,
Shanghai Jiao Tong University, Shanghai, China}
\affiliation{${}^{3}$CTS, CASTS and
Department of Physics, National Taiwan University, Taipei, Taiwan\\}
\affiliation{${}^{4}$National Center for
Theoretical Sciences and Physics Department of National Tsing Hua University, Hsinchu, Taiwan}

\date{\today $\vphantom{\bigg|_{\bigg|}^|}$}

\date{\today}

\vskip 1cm
\begin{abstract}
We study a class of models for neutrino mass matrix in Type-II seesaw with $A_4$ family symmetry.
The resulting neutrino mass matrix can be naturally made to respect a $\mu-\tau$ exchange plus CP conjugate symmetry (GLS) with the CP violating phase $\delta$ and the mixing angle $\theta_{23}$  predicted to be $\pm\pi/2$ and $\pi/4$, respectively. When GLS is explicitly broken by complex Yukawa couplings, the model predictions for $\delta$ and $\theta_{23}$ can be significantly modified.  Should future experiments will indeed determine $\theta_{23}$ and $\delta_{CP}$ away from the GLS limit values, one then had to consider models with broken GLS. We study several simple scenarios to show how the modifications arise  when GLS is broken and how future experiments can test this class of models.
\end{abstract}

\pacs{PACS numbers: }

\maketitle

\section{Introduction}

Experiments have collected a lot of precious information about the neutrino mixing parameters. The mixing angles in the Pontecorvo-Maki-Nakagawa-Sakata  matrix\cite{pmns} $V_{PMNS}$ are not always small \cite{pdg,valle,fogli-schwetz} as their quark mixing counter part\cite{pdg,km}. In the standard parametrization\cite{pdg, ckm} for three neutrino mixing commonly used\cite{valle,fogli-schwetz}, the mixing angle $\theta_{23}$ is close to $\pi/4$, $\theta_{12}$ is also large, $\theta_{13}$ is relatively small but away from zero. There are also evidences showing that the CP violating Dirac phase $\delta$ is close to $-\pi/2$ (or $3\pi/2$). If these data are further confirmed, the neutrino mass matrix will have a simple form. Assuming that neutrinos are Majorana particles,  the neutrino mass matrix defined by the term giving neutrino masses in the Lagrangian
$(1/2)\bar \nu_L m_\nu \nu^c_L$ has the following form,
\begin{eqnarray}
m_\nu = V_{PMNS} \hat m_\nu V^T_{PMNS}\;,
\end{eqnarray}
where $\hat m_\nu = diag(m_1, m_2, m_3)$ with $m_i = |m_i|exp(i\alpha_i)$. Here we have put Majorana phase information in the neutrino masses. The standard form for $V_{PMNS}$ is given by
\begin{eqnarray}
V_{PMNS} = U(\theta_{12}, \theta_{13},\theta_{23}, \delta) =  \left(\!\begin{array}{ccc}
 c_{12\,}^{}c_{13}^{} & s_{12\,}^{}c_{13}^{} & s_{13}^{}\,e^{-i\delta}
\vspace{1pt} \\
-s_{12\,}^{}c_{23}^{}-c_{12\,}^{}s_{23\,}^{}s_{13}^{}\,e^{i\delta} & ~~
 c_{12\,}^{}c_{23}^{}-s_{12\,}^{}s_{23\,}^{}s_{13}^{}\,e^{i\delta} ~~ & s_{23\,}^{}c_{13}^{}
\vspace{1pt} \\
 s_{12\,}^{}s_{23}^{}-c_{12\,}^{}c_{23\,}^{}s_{13}^{}\,e^{i\delta} &
-c_{12\,}^{}s_{23}^{}-s_{12\,}^{}c_{23\,}^{}s_{13}^{}\,e^{i\delta} & c_{23\,}^{}c_{13}^{}
\end{array}\right) ,
\label{pmns}
\end{eqnarray}
where $c_{ij}$ and $s_{ij}$ are $\cos\theta_{ij}$ and $\sin\theta_{ij}$, respectively. They are all normalized to be positive.

With $\delta = -\pi/2$ and $\theta_{23}=\pi/4$, $m_\nu$ has the following form\cite{ma-a4,grimus,xing}
\begin{eqnarray}
m_\nu = \left (\begin{array}{ccc}
a& c+i\beta&-(c - i\beta)\\
c+i\beta&d+i\gamma&\tilde b\\
-(c-i\beta)&\tilde b&d-i\gamma
\end{array}
\right )\;,
\end{eqnarray}
where
\begin{eqnarray}
&&a=m_1c^2_{12} c^2_{13}+m_2s^2_{12} c^2_{13}-m_3 s^2_{13}\;,\;\;\;\;\;\;\;\;\;\tilde b =-\frac{1}{2} \left(m_1(s^2_{12} + c^2_{12} s^2_{13}) +m_2(c^2_{12}+ s^2_{12}s^2_{13})-m_3 c^2_{13}\right)\;,\nonumber\\
&&c=- \frac{1}{\sqrt{2}} (m_1-m_2) s_{12}c_{12} c_{13}\;,\;\;\;\;\;\;\;\;\;\;\;\;\;\;\;
d=\frac{1}{2} \left(m_1 (s^2_{12} - c^2_{12} s^2_{13})
+m_2 (c^2_{12} - s^2_{12} s^2_{13})+m_3 c^2_{13}\right )\;,\nonumber\\
&&\beta = \frac{1}{\sqrt{2}} s_{13} c_{13}\left(m_1 c_{12}^2+m_2 s^2_{12}+m_3\right)\;,\;\;\gamma =   -  (m_1-m_2) s_{12} c_{12} s_{13} \;.
\label{elements}
\end{eqnarray}
Note that in the most general case, because non-zero Majorana phases, the parameters $a$, $\tilde b$, $c$, $d$, $\beta$ and $\gamma$ are all complex.

One has the degrees of freedom to redefine the neutrino fields phases and the most general form of the above mass matrix can be rewritten as
\begin{eqnarray}
m_\nu = \left (\begin{array}{ccc}
e^{ip_1}&0&0\\
0&e^{ip_2}&0\\
0&0&e^{ip_3}
\end{array}\right )
\left (\begin{array}{ccc}
a& c+i\beta&-(c - i\beta)\\
c+i\beta&d+i\gamma&\tilde b\\
-(c-i\beta)&\tilde b&d-i\gamma
\end{array}
\right )\left (\begin{array}{ccc}
e^{ip_1}&0&0\\
0&e^{ip_2}&0\\
0&0&e^{ip_3}
\end{array}\right )\;,
\end{eqnarray}
where the phases $p_i$ are arbitrary. One can choose some particular values for $p_i$ to obtain forms of $m_\nu$ for convenience of analysis. For example the ``-'' sign for the ``13'' and ``31'' entries can be removed by choosing
$p_1=p_2=0$ and $p_3 =\pi$, the resultant matrix can be written in a more familiar forms
\begin{eqnarray}
m_\nu = \left (\begin{array}{ccc}
a& c+i\beta&(c - i\beta)\\
c+i\beta&d+i\gamma&b\\
(c-i\beta)&b&d-i\gamma
\end{array}
\right ), \label{matrix}
\end{eqnarray}
where $b = - \tilde b$.

If neutrinos do not have non-trivial Majorana phases (but mass can be positive or negative), the mass matrix has the following form
\begin{eqnarray}
 m_\nu &=&
\left (\begin{array}{ccc}
A& C& C^*\\
 C& D^*& B\\
 C^*& B& D
\end{array}
\right )\;.\label{real}
\end{eqnarray}

Replacing $\delta = -\pi/2$ by $\delta = \pi/2$, the neutrino mass matrix is given in a similar form as that in eq.(\ref{matrix}), but $\beta$ and $\gamma$ need to be multiplied by a ``-'' sign. This implies that without further information given, a general mass matrix in the form given by eq.(\ref{matrix}) can give $\delta = \pm \pi/2$ and $\theta_{23}=\pi/4$. Whether they predict $+\pi/2$ or $-\pi/2$, additional information need to be provided\cite{he-new}.
It has been pointed out that the general form in eq.(\ref{matrix}) is a necessary condition for $\delta = \pm\pi/2$ and $\theta_{23} = \pi/4$, but not sufficient condition. In our later discussions, unless specified, the mass matrix of the form in eq.(\ref{matrix}) is always referred to a general form whose elements are not necessarily given by those in eq.(\ref{elements}). While the mass matrix in the form of eq.(\ref{real}) provide sufficient condition for $\delta=\pm \pi/2$ and $\theta_{23}=\pi/4$ when $s_{13}$ and $\sin\delta$ are not zero. The simplicity of the above mass matrix may serve as a good starting point to understand possible underlying theory.  In fact  it has been shown  that the above
neutrino mass matrix is a consequence of imposing a symmetry of the form $e \to e$, $\mu$ and $\tau$ exchange with $CP$ conjugation discussed by Grimus and Lavoura in Ref.\cite{grimus0}, which we will refer to as the Grimus-Lavoura symmetry (GLS).

In this work, we study realizations of $\delta=\pm \pi/2$ and $\theta_{23} = \pi/4$ in type-II seesaw model with $A_4$ flavor symmetry. Models based on $A_4$ symmetry has been shown to be able to provide a good scenario to achieve this\cite{he-new,ma-new}. In $A_4$ models, the charged lepton mass matrix $M_l$ is diagonalized from left (rotation on left-handed charged leptons) by the characteristic matrix $U_l$ for $A_4$ symmetry model buildings\cite{ma-a4},
\begin{eqnarray}
M_l = U_l \hat m_l U_r\;,\;\;
U_l = {1\over \sqrt{3}}\left ( \begin{array}{lll}
1&1&1\\
1&\omega&\omega^2\\
1&\omega^2&\omega
\end{array}\right )\;,\label{ul}
\end{eqnarray}
where $\omega = exp(i 2\pi/3)$ and $\omega^2 = exp(i4\pi/3)$.
$U_r$ is a unitary matrix, but does not play a role in determining $V_{PMNS}$. We will not specify its form here.

If neutrinos are Majorana particles, the most general mass matrix is of the form
\begin{eqnarray}
M_\nu =  \left ( \begin{array}{lll}
w_1&x&y\\x&w_2&z\\y&z&w_3
\end{array}\right )\;, \label{g-matrix}
\end{eqnarray}
which can be diagonalized by unitary matrix $V_\nu$, $M_\nu = V_\nu \hat m_\nu V_\nu^T$.

The mixing matrix $V_{PMNS}$ is given by
\begin{eqnarray}
V_{PMNS} = U^\dagger_l V_\nu\;. \label{mixing-define}
\end{eqnarray}
In the basis where charged lepton is diagonalized, the neutrino mass matrix is of the form given by eq.(\ref{matrix})  with\cite{ma-new}
\begin{eqnarray}
&&a ={1\over 3}(w_1+w_2+w_3+2(x+y+z))\;,\nonumber\\
&&d+i\gamma={1\over 3}(w_1 + \omega w_2 + \omega^2 w_3+ 2(\omega^2 x+\omega y+z))\;,\nonumber\\
&&d-i\gamma ={1\over 3}(w_1 + \omega^2 w_2 + \omega w_3+2(\omega x+\omega^2 y+z))\;,\\
&&c+i\beta={1\over 3}(w_1+\omega^2 w_2 + \omega w_3-\omega x-\omega^2 y-z)\;,\nonumber\\
&&c-i\beta={1\over 3}(w_1 + \omega w_2 + \omega^2 w_3-\omega^2 x-\omega y-z)\;,\nonumber\\
&&b={1\over 3}(w_1+w_2+w_3- (x+y+z))\;.\nonumber
\label{model-matrix}
\end{eqnarray}

 If one imposes the GLS on the neutrino mass matrix,  all parameters in the set $P = (w_i, x, y, z)$ are dictated to be real, and will predict\cite{he-new,ma-new} $\delta = \pm \pi/2$ and $\theta_{23} = \pi/4$. Therefore in $A_4$ model building for neutrino masses with $\delta=\pm \pi/2$ and $\theta_{23} = \pi/4$,  it is essentially to make sure that $U_l$ is of the form given by eq.(\ref{ul}) and require the resulting mass matrix to satisfy GLS. Note that in this case since the parameters in set $P$ are all real, the complexity of the mixing matrix is purely due to the appearance of $\omega$ and $\omega^2$.
When the GLS is broken by allowing the parameters in $P$ can be complex, there are more sources for CP violation and the model does not predict $\delta = \pm \pi/2$ and $\theta_{23} = \pi/4$ automatically. This points a way to modify the predictions to fit data should future experiments will find $\delta$ and $\theta_{23}$ to be deviate significantly from $-\pi/2$ and $\pi/4$. We will study both cases with the parameters in set $P$ to be real and complex  in the rest of the paper.

\section {Type-II seesaw model with $A_4$ symmetry}

We now construct a Type-II seesaw model\cite{seesaw2} with $A_4$ family symmetry to realize the forms of mass matrices in eqs.(\ref{matrix}) and (\ref{real}).  A different model based on Type-II seesaw with $A_4$ has been constructed to realized tribi-maximal neutrino mixing\cite{sugiyama}. In our model, the left-handed lepton doublet $l_L$ and right-handed charged lepton singlet $l_R$ have the following standard $SU(3)_C\times SU(2)_L\times U(1)_Y$ gauge and $A_4$ family symmetry properties
\begin{eqnarray}
l_L: (1, 2, -1)(3)\;,\;\; l_R : (1,1,-2)(1+1''+1')\;,
\end{eqnarray}
where the first three numbers in the first bracket indicate the $SU(3)_C$,  $SU(2)_L$ and $U(1)_Y$ transformation properties. The numbers in the second bracket indicate the $A_4$ representations.

To obtain desired mixing pattern, the Higgs sector is enlarged to have two types of Higgs
doublets, $\phi$ and $\Phi$, and two triplets, $\Delta$ and $\chi$ for neutrino masses. They transform as
\begin{eqnarray}
\phi: (1, 2, -1)(1)\;,\;\; \Phi : (1, 2, -1)(3)\;,\;\;\Delta^{0,',''}: (1, 3, -2)(1+1'+1'')\;,\;\;\chi: (1, 3, -2)(3)\;.
\end{eqnarray}

The Lagrangian responsible for the lepton mass matrix is
\begin{eqnarray}
L &=& y_e \bar l_L \tilde \Phi l^1_R  + y_\mu \bar l_L
\tilde \Phi l^2_R + y_\tau \bar l_L \tilde \Phi
l^3_R\nonumber\\
& +& Y^0_\nu \bar l_L \Delta^0 l^c_L +Y^{'}_\nu \bar l_L \Delta^{'} l^c_L+Y^{''}_\nu \bar l_L \Delta^{''} l^c_L+y_\nu \bar \l_L \chi l^c_L +H.C.
\end{eqnarray}

If the structure of the vacuum expectation value (vev) is of the form $<\Phi_{i}> = v^{\Phi}_i$,
$<\chi_{i}>=v^\chi_{i}$, $<\phi> = v_\phi$, and $<\Delta^{0,',''}> = v^{0,',''}_\Delta$, one
 obtains the charged lepton and neutrino mass matrices $M_l$  and $M_\nu$ as
\begin{eqnarray}
M_l=
\sqrt{3}\left ( \begin{array}{lll}
v^\Phi_1&0&0\\0&v^\Phi_2&0\\0&0&
v^\Phi_3\end{array}\right )U_l \left ( \begin{array}{lll}
y_e &0&0\\0&y_\mu &0\\0&0&y_\tau
\end{array}\right )\;, \;\;M_\nu =
\left ( \begin{array}{lll}
w_1&x&y\\x&w_2&z\\y&z&w_3
\end{array}\right )\;, \label{m-matrix}
\end{eqnarray}
where
\begin{eqnarray}
&&w_1=Y^0_\nu v_\Delta^0+Y^{'}_\nu v_\Delta^{'}+Y^{''}_\nu v_\Delta^{''}\;,\nonumber\\
&&w_2=Y^0_\nu v_\Delta^0+\omega^2 Y^{'}_\nu v_\Delta^{'}+\omega Y^{''}_\nu v_\Delta^{''}\;,\nonumber\\
&&w_3=Y^0_\nu v_\Delta^0+\omega Y^{'}_\nu v_\Delta^{'}+\omega^2 Y^{''}_\nu v_\Delta^{''}\;,\nonumber\\
&&x=y_\nu v^\chi_3\;,\;\;y=y_\nu v^\chi_2\;,\;\;z=y_\nu v^\chi_1\;. \label{eqss}
\end{eqnarray}

If the vevs of $<\Phi_{1,2,3}> = v^\Phi_{1,2,3}$ are all equal to  $v^\Phi$, the vev structure of the Higgs fields breaks $A_4$, but left with a $Z_3$ residual symmetry generated by
$\{I,c,a\}$. Here $c,\;a$ are $A_4$ group elements defined
in Ref.\cite{hev}. This will lead to the charged lepton mass matrix of the form
\begin{eqnarray}
M_l =
U_l \left ( \begin{array}{lll}
m_e &0&0\\0&m_\mu&0\\0&0&m_\tau
\end{array}\right )\;,
\end{eqnarray}
where $m_{e,\mu,\tau} = \sqrt{3}y_{e,\mu,\tau} v^\Phi$.

As long as $Z_3$ residual symmetry in the model is not broken, we have one of the key ingredients in realizing the form of charged lepton mass matrix in eq.(\ref{ul}). In the basis where the charged lepton mass matrix is diagonalized, the neutrino mass matrix will be the same form in eq.(\ref{matrix}).

In the rest of this paper, we will study consequences related to the above mass matrix.  We will study conditions on the model parameters imposed by GLS, and phenomenologically acceptable models can result in this class of models.

Before carrying out detailed analysis for neutrino mixing, we would like to point out that this model can easily accommodate data in the quark sector if one assigns the left- and right- handed quarks $Q_L$, $U_R$ ad $D_R$
as $A_4$ singlet ``1''. In this case the Yukawa couplings for quark masses and their mixing are given by
\begin{eqnarray}
L = \bar Q_L Y_u \phi U_R + \bar Y_d \bar Q_L \tilde \phi D_R + H.C.\;.
\end{eqnarray}
This will give, in general, arbitrary $3\times 3$ up- and down- quark mass matrices $M_{u,d}$ after $\phi$ develops a non-zero vev $v_\phi$ with  $M_{u,d} = Y_{u,d} v_\phi$. These matrices will have no predictive power for quark masses and their mixing, but have no problem in accommodating experimental data.

\section{Mass matrix predicting $\delta =\pm \pi/2$ and $\theta_{23}=\pi/4$}

If the parameters in the set $P$ are all real, the resulting mass matrix is of the form given in eq.(\ref{real}). Therefore this model predicts
\begin{eqnarray}
\delta = \pm {\pi\over 2}\;,\;\;\theta_{23} = {\pi\over 4}\;.
\end{eqnarray}.

The above prediction can also be obtained by studying the mixing matrix $V_{PMNS}$ is eq.(\ref{mixing-define}).
For an arbitrary complex symmetric matrix $M_\nu$, the matrix $V_\nu$ diagonalizes it, has the most general form
\begin{eqnarray}
V_\nu = diag(1, e^{i\tau_2}, e^{i\tau_3})U(\tilde \theta_{12}, \tilde \theta_{13}, \tilde \theta_{23}, \tilde \delta)\;.
\end{eqnarray}
Here we have absorbed a Majorana phase matrix $diag(1,e^{i\eta_2}, e^{i\eta_3})$ on the right of the above equation into the eigen-masses $m_i e^{i2\eta_i}$.

If the parameter set in $P$ are all real, the mass matrix $M_\nu$ can be diagonalized by an orthogonal mixing matrix
$V_\nu=U(\tilde \theta_{12}, \tilde \theta_{13}, \tilde \theta_{23}, \tilde \delta =0)$,  and $e^{i\tau_i}$ are real. Depending on conventions, here $\tilde s_{ij}$ and $\tilde c_{ij}$ need not to be normalized all to be positive. We have the  $V_{PMNS}$ given as the following
\begin{eqnarray}
\left ( \begin{array}{ccc}
{-\tilde s_{12} (\tilde c_{23} - \tilde s_{23}) +\tilde c_{12} (\tilde c_{13} - \tilde s_{13} (\tilde c_{23} +\tilde s_{23}))\over \sqrt{3}}
&{\tilde c_{12} (\tilde c_{23} -\tilde s_{23}) +\tilde s_{12} (\tilde c_{13} - \tilde s_{13} (\tilde c_{23} +\tilde s_{23}))\over \sqrt{3}}
&{\tilde s_{13} +\tilde c_{13} (\tilde c_{23} +\tilde s_{23})\over \sqrt{3}}\\
{-\tilde s_{12}(\omega^2\tilde c_{23}  - \omega\tilde s_{23}) +\tilde c_{12} (\tilde c_{13} - \tilde s_{13} (\omega \tilde c_{23}  + \omega^2\tilde s_{23}))\over \sqrt{3}}
&{\tilde c_{12} (\omega^2\tilde c_{23}  - \omega\tilde s_{23}) +\tilde s_{12} (\tilde c_{13} - \tilde s_{13} (\omega \tilde c_{23}  + \omega^2\tilde s_{23}))\over \sqrt{3}}
& {\tilde s_{13} +\tilde c_{13}(\omega \tilde c_{23}  + \omega^2 \tilde s_{23})\over \sqrt{3}}\\
{-\tilde s_{12}(\omega\tilde c_{23} - \omega^2 \tilde s_{23}) +\tilde c_{12} (\tilde c_{13} - \tilde s_{13} (\omega^2 \tilde c_{23} +\omega \tilde s_{23}))\over \sqrt{3}}
&{\tilde c_{12} (\omega \tilde c_{23} -\omega^2\tilde s_{23}) +\tilde s_{12} (\tilde c_{13} - \tilde s_{13} (\omega^2 \tilde c_{23} +\omega\tilde s_{23}))\over \sqrt{3}}
&{\tilde s_{13} +\tilde c_{13}(\omega^2\tilde c_{23} +  \omega\tilde s_{23})\over \sqrt{3}}
\end{array} \right ).
\end{eqnarray}

The above matrix has the property that, $|V_{PMNS,2i}| = |V_{PMNS, 3i}|$. Rewriting the above into the standard form of the mixing matrix, this leads to\cite{grimus}
\begin{eqnarray}
\theta_{23} = {\pi\over 4}\;, \;\;s_{13}\cos\delta =0.
\end{eqnarray}
Since experimentally $s_{13}\neq 0$, then $\cos\delta =0$ which leads to $\delta = \pm \pi/2$.

The other angles can be determined by
\begin{eqnarray}
s_{13} = |V_{13}|\;,\;\;t_{12} = {|V_{12}|\over |V_{11}|}\;, \label{angles}
\end{eqnarray}
where $t_{ij} = s_{ij}/c_{ij}$.

The condition whether $\delta$ takes $+\pi/2$ or $-\pi/2$ is determined by using the Jarlskog parameter\cite{jarlskog}
$J = c^2_{13}s_{12}c_{12}s_{23}c_{23}s_{13}\sin\delta = Im(V_{11}V_{12}^*V_{21}^*V_{22})$ to obtain
\begin{eqnarray}
\sin\delta = {(1-|V_{13}|^2)Im(V_{11}V_{12}^*V_{21}^*V_{22})\over |V_{11}||V_{12}||V_{23}||V_{33}||V_{13}|}\;.\label{j-parameter}
\end{eqnarray}

Note that if $\tilde \delta \neq 0$ and/or $e^{i\tau_i}$ are not real, in general $|V_{PMNS, 2i}|$ is not the same as
$|V_{PMNS, 3i}|$. $\theta_{23}$ and $\delta$ will not necessarily be $\pi/4$ and $\pm \pi/2$. One should be cautious about this. This uncertainty actually can be turned into a good use to provide a possible way to modify the predictions for $\theta_{23}$ and $\delta$.
The value for $\theta_{23}$ is determined by
\begin{eqnarray}
t_{23} = {|V_{23}|\over |V_{3 3}|}\;,\label{angle23}
\end{eqnarray}
The angles $\theta_{13}$ and $\theta_{12}$ are determined by eq.(\ref{angles}), and the CP violating phase $\delta$ is given by eq.(\ref{j-parameter}).

 The above discussions provides a way of looking at how  the reality of the parameters in the set $P$ leads to $\delta = \pm \pi/2$ and $\theta_{23}=\pi/4$. It is essential to have the parameters in the set $P$ to be all real. The complexity of the parameters can appear in the Yukawa couplings, in the vevs, and also in places where $\omega^i$ appear in neutrino mass matrix. One needs to see with what conditions they can be made real.

To make the Yukawa couplings and scalar vevs real, one can require the model Lagrangian to satisfy a generalized CP symmetry under which
\begin{eqnarray}
&&(l_{e,L} \;,\;\;l_{\mu,L}\;,\;l_{\tau,L})  \to ( l_{e,L}^{CP}\;,\;\; l_{\tau,L}^{CP}\;,\;\;l_{\mu,L}^{CP})\;,\;\;\;\Phi = (\Phi_1, \Phi_2, \Phi_3) \to (\Phi^\dagger_1, \Phi^\dagger_3, \Phi^\dagger_2)\;,\\
&&(\Delta^0\;,\;\Delta^{'}\;,\;\;\Delta^{''})  \to (\Delta^{0\dagger}\;, \;\;\Delta^{'\dagger}\;,\;\; \Delta^{''\dagger})\;,\;\;\;\;(\chi_1\;,\;\chi_2\;,\;\;\chi_3)  \to (\chi_1^\dagger\;, \;\;\chi_3^{\dagger}\;,\;\; \chi_2^\dagger)\;,\nonumber
\end{eqnarray}
and all other fields transform  the same as those under the usual CP symmetry. Here the superscript $CP$ in the above indicate that the fields are the usual $CP$ transformed fields.

The above transformation properties will transform relevant terms into their complex conjugate ones. Requiring the Lagrangian to be invariant under the above transformation dictates the Yukawa couplings to be real. The same requirement will dictates the scalar potential to forbid spontaneous CP violation and vevs to be real. One, however, notices that the parameters
$w_{2,3}$ in eq.(\ref{eqss}) are in general complex even if the Yukawa couplings and the vevs of the scalar fields are made real because the appearance of $\omega^i$. To make them real to satisfy GLS, it is therefore required that
\begin{eqnarray}
Im(\omega^2 Y^{'}_\nu v_\Delta^{'}+\omega Y^{''}_\nu v_\Delta^{''})=Im(\omega Y^{'}_\nu v_\Delta^{'}+\omega^2 Y^{''}_\nu v_\Delta^{''}) = 0\;.
\end{eqnarray}
The above can be achieved by the absent of the scalar fields $\Delta^{',''}$ in the theory or $Y^{'}_\nu v_{\Delta'}= Y^{''}_\nu v_{\Delta''}$.  We will consider examples for each case in the next section.

 From the above discussions, we know that in general the mass matrices obtained can accommodate values different than $\pi/4$ and $\pm\pi/2$ for $\theta_{23}$ and $\delta$ respectively, we will not carry out a full numerical search analysis for parameter spaces, but to take some simple cases to show how the modifications arise when GLS is broken and how future experiments can test this class of models. These models, due to additional constraints, will have some additional predictions than the general one.

\subsection{Model A: $Z_2$ residual symmetry for neutrino mass matrix}

If vev of $\chi_2$ component of $\chi$ is
non-zero, but the vevs of $\chi_{1,3}$ are zero, the vev structure breaks $A_4$ down to a $Z_2$ generated by
$\{1,r_2\}$. Here $r_2$ is an $A_4$ group element defined
in Ref.\cite{hev}.  The charge lepton and neutrino mass matrices are given by
\begin{eqnarray}
M_l =
U_l \left ( \begin{array}{lll}
m_e &0&0\\0&m_\mu&0\\0&0&m_\tau
\end{array}\right )\;,\;\;\;\;
M_{\nu }=\left(
\begin{array}{lll}
w_1 & 0 & y  \\
0 & w_2  & 0 \\
y  & 0 & w_3
\end{array}
\right)\;.
\end{eqnarray}
The parameters in the set $w_i$, and $y$ are in general complex.

If the fields, $\Delta^{'}$ and $\Delta^{''}$ are absent from the model, one would have $w_1=w_2=w_3 = Y^0_\nu v_\Delta^0$. The resulting mass matrices lead to the well known tribi-maximal mixing\cite{tribi} which had been the focus for $A_4$ symmetry model buildings before\cite{hev,alterali-hebabu,zee,he1}. In this case, $\theta_{13}=0$ and phase $\delta$ is zero.  Since experimentally $s_{13}$ has been determined to be non-zero, this model is ruled out. Modifications have to be implemented\cite{hezee, hezee1,hezee2} to accommodate data. A way out is to keep $Y^{'}_\nu v_{\Delta'}=Y^{''}_\nu v_{\Delta^{''}}$ in the model as discussed in the previous section.

We now summarize the main results follow Ref.\cite{he-new} for this model. This also services as an outline how analysis can be carried out for relevant models.
Letting $M_\nu$ be diagonalized by $\tilde V_\nu$, one has
\begin{eqnarray}
M_\nu = \tilde V_\nu \hat m_\nu \tilde V_\nu^T\;. \label{mmnu}
\end{eqnarray}
$\tilde V_\nu$ in this case can always be written in the following way with Majorana phases to be absorbed in to the neutrino masses $m_i$,
\begin{eqnarray}
\tilde V_\nu = V_\rho V_\nu\;,\;\; \mbox{with}\;\;V_\nu = \left (\begin{array}{ccc}
c&0&-s\\
0&1&0\\
s&0&c
\end{array}
\right )\;,
\end{eqnarray}
where $V_\rho$ is a diagonal matrix $diag(1,1,e^{i\rho})$ with $\tan\rho = Im (yw_1^*+y^*w_3)/Re(yw_1^*+y^*w_3)$.
$s = \sin\theta$ and $c = \cos\theta$.

Expressing the mixing angle $\theta$ in terms of the model parameters, we obtain
\begin{eqnarray}
\tan2\theta = {2|yw^*_1+w_3 y^*|\over |w_1|^2 - |w_3|^2}\;.
\end{eqnarray}

Using $\hat m_\nu = V^\dagger_\nu V^\dagger_\rho M_\nu V^*_\rho V^*_\nu$ derived from eq.(\ref{mmnu}), we obtain the Majorana phases $\alpha_i$ of $m_i$ in terms of the model parameters as
\begin{eqnarray}
\alpha_{1,3} = Arg(w_i(1\pm \cos2\theta) + w_2 e^{-i2\rho}(1\mp \cos2\theta) \pm 2 \sin2\theta y e^{-i\rho}\;,\;\;\alpha_2 = Arg(w_2)\;.
\end{eqnarray}
One can always normalize the Majorana phase $\alpha_2$ to be zero without of generality. In this basis, the phase of $w_2$ is also zero. Although the Majorana phases can be expressed in terms of the model parameters, since there are more number of parameters than the mixing angles and eigen-masses, there is no prediction for the Majorana phases.
We will not discuss them in our later numerical analysis any more.

The mixing matrix $V_{PMNS}$ can be, in general, written as
\begin{eqnarray}
V_{PMNS} = U_l^\dagger V_\rho V_\nu =  {1\over \sqrt{3}}\left(
\begin{array}{ccc}
c + se^{i\rho} & 1 &
ce^{i\rho}-s \\
c+\omega se^{i\rho}& \omega^2  &
\omega ce^{i\rho} - s \\
c+\omega^2 se^{i\rho}& \omega & \omega^2 ce^{i\rho} -s
\end{array}
\right)\;,
\end{eqnarray}

Using eqs.(\ref{angles}), (\ref{j-parameter}) and (\ref{angle23}),
we find that
\begin{eqnarray}
s_{12} = {1\over \sqrt{2}(1+cs\cos\rho)^{1/2}}\;,\;s_{23} = {(1+cs \cos\rho + \sqrt{3} cs \sin\rho)^{1/2}\over \sqrt{2}(1+cs \cos\rho)^{1\over 2}}\;,\;s_{13} = {(1-2cs \cos \rho)^{1/2}\over \sqrt{3}}\;.
\end{eqnarray}
and
\begin{eqnarray}
\sin\delta = (1+{4 c^2s^2 \sin^2\rho\over (c^2-s^2)^2})^{-1/2}(1- {3 c^2s^2\sin^2\rho\over (1+cs \cos\rho)^2})^{-1/2} \times
\left \{ \begin{array} {l}
-1\;,\;\; \mbox{if}\;\;c^2>s^2\;,\\
+1\;,\;\;\mbox{if}\;\;s^2>c^2\;.
\end{array} \right .
\end{eqnarray}
From the above, one clearly sees that if $\sin\rho$ is not zero, $|\delta|$ and $\theta_{23}$ deviate from $\pi/2$ and $\pi/4$, respectively. In the limit $\rho$ goes to zero, however,  the above recovers the results with real parameter set $P$ with $\delta = \pm \pi/2$ and $\theta_{23} = \pi/4$.

There are two interesting features for this model worth mentioning. One of is that $|V_{e2}|$ to be $1/\sqrt{3}$ which agree with date. $s_{12}$ is always larger or equal to $1/\sqrt{3}$ which is a decisive test for this model.
Another is that although the Dirac phase $\delta$ depends on the phase $\rho$, the Jarlskog parameter $J$ which is independent of $\rho$ given by $J = -(c^2-s^2)/6\sqrt{3}$. This implies that CP violation related to neutrino oscillation is still purely due to intrinsic CP violation. This model can be made in agreement with data at 1$\sigma$ level.

If $\rho=0$ and $c=s=1/\sqrt{2}$, the mixing pattern is the tribi-maximal. However, if $\rho$ is not zero, even if
$c=s=1/\sqrt{2}$, $s_{13}$ can be non-zero,$s_{12}$ and $s_{23}$ are also modified from their tribi-maximal values
\begin{eqnarray}
s_{12} = {1\over (2+\cos\rho)^{1/2}}\;,\;s_{23} = {1\over \sqrt{2}}(1 + {\sqrt{3} \sin\rho\over 2+ \cos\rho})^{1/ 2}\;,\;s_{13} = {(1-\cos \rho)^{1/2}\over \sqrt{3}}\;.\label{special}
\end{eqnarray}
$J$ is exactly zero which implies $\sin\delta =0$.


\subsection{Model B: mass matrix with $w_{1,2,3}=0$}

Without $\Delta^{0,',''}$, $w_{1,2,3} = 0$. In this case all the three vevs of $\chi$ should not be equal in order to fit data on the neutrino mass-squared differences $\Delta m^2_{21} = |m_2|^2- |m_1|^2$ and $\Delta m^2_{31} = |m_3|^2- |m_1|^2$. The $Z_2$ residual symmetry mentioned previously is also broken in this case. If CP is spontaneously broken in the Higgs potential, $v^\chi_i$ may be complex too. Therefore the parameters in the set $P: \{x, y, z \} = \{|x|e^{i\sigma_x}, |y|e^{i\sigma_y}, |z|e^{i\sigma_z}\}$ are in general complex.
One can always choose, without loss of generality, to rewrite the neutrino mass matrix as $M_\nu = V_\sigma \tilde M_\nu V_\sigma^T$
\begin{eqnarray}
\tilde M_\nu = \left (\begin{array}{lll}
0&|x|&|y|\\
|x|&0&|z|\\
|y|&|z|&0
\end{array}
\right )\;,\;\;
V_\sigma = e^{i\sigma_1}\left (\begin{array}{lll}
1&0&0\\
0&e^{i\sigma_2}&0\\
0&0&e^{i\sigma_3}
\end{array}
\right )\;,
\end{eqnarray}
where $\sigma_1 = (\sigma_z-\sigma_x-\sigma_y)/2$, $\sigma_2=\sigma_z-\sigma_y$ and $\sigma_3=\sigma_x-\sigma_y$. $\sigma_1$ can be absorbed into lepton fields. We will neglect it in the following analysis.

Let us discuss now how the model parameters can be determined by data.
First of all since $\tilde M_\nu$ is a real matrix, the eigen-values $\tilde m_i$ are all real and $\tilde m_1 + \tilde m_2+\tilde m_3 = 0$. The absolute value of neutrino mass $|m_i|=|\tilde m_i|$ can be solved as functions of $\Delta m^2_{21}$ and $\Delta m^2_{31}$. For the values allowed by experimental data, the solutions are given by up to a overall sign
\begin{eqnarray}
&&\tilde m_{1} = \frac{(\Delta m_{21}^2-\Delta X^2)\sqrt{\Delta m_{21}^2-\Delta m_{31}^2+2 \Delta X^2} }{\sqrt{3} (\Delta m_{21}^2+\Delta m_{31}^2)}\nonumber\\
&&\tilde m_{3} = \frac{(\Delta m_{31}^2+\Delta X^2)\sqrt{\Delta m_{21}^2-\Delta m_{31}^2+2 \Delta X^2} }{\sqrt{3} (\Delta m_{21}^2+\Delta m_{31}^2)}\nonumber\\
&&\Delta X^2  =  \sqrt{(\Delta m_{21}^2)^2+\Delta m_{21}^2 \Delta m_{31}^2+(\Delta m_{31}^2)^2}
\end{eqnarray}
One can easily check that multiplying a ``$-$'' to the above one obtains another  solution.   The overall sign can be absorbed into lepton field phases redefinition.
 We will use the above normalization for the overall sign.
In the numerical study later, we will show that only inverted hierarchy is allowed. Therefore, $m_{1,3}>0$ and
$m_2 <0$. Since the eigen-masses are all real, there is no Majorana phases in this model.

Since $\tilde M_\nu$ is real, it can be diagonalized by a orthogonal matrix. Assuming $U(\tilde \theta_{13}, \tilde \theta_{13}, \tilde \theta_{23}, \tilde \delta = 0)$ diagonalizes $\tilde M_\nu$, one can express the fact that the ``11''' and ``22'' entries are zero as
\begin{eqnarray}
&&\tilde c_{13}^2 (\tilde m_1 \tilde c_{12}^2 + \tilde m_2 \tilde s_{12}^2) +\tilde m_3 \tilde s_{13}^2 =0\;,\nonumber\\
&&\nonumber\\
&&\tilde m_1 (  \tilde s_{13} - \tilde t_{12} \tilde t_{23}) ( \tilde t_{12}   +  \tilde t_{23}\tilde  s_{13} ) -
 \tilde m_2 (\tilde t_{12} \tilde  s_{13} +  \tilde t_{23}) (1 - ‪\tilde t_{12} \tilde t_{23} \tilde s_{13} ) + \tilde m_3  {\tilde c_{13}^2\over \tilde c_{12} }\tilde t_{23} =0\;, \label{model-b}
\end{eqnarray}
where $\tilde t_{ij} = \tilde s_{ij}/\tilde c_{ij}$.

From the above two equations, one can express $\tilde s_{12} $ and $\tilde s_{23}$ as functions of $\tilde s_{13}$ and $\tilde m_i$. Since $\tilde m_i$ are known, the elements in $V_{PMNS}=U^\dagger_l V_\sigma V_\nu$ can be expressed as functions of $\sigma_2$, $\sigma_3$ and $\tilde s_{13}$. Using the relations between the elements $V_{ij}$ of $V_{PMNS}$ with $s_{ij}$ and $\delta$ in eqs.(\ref{angles}) and (\ref{j-parameter}),
one can vary $\sigma_2$, $\sigma_3$ and $\tilde s_{13}$ to see if the resulting values for $s_{12, 13, 23}$ are in the allowed values, and obtain $\delta$ by eq.(\ref{j-parameter}).

Similar analysis had been used to rule out\cite{rule-out-z-model} the simple version of Zee model where the mass matrix for neutrinos is $\tilde M_\nu$ in the diagonalized basis of the charged leptons. Here the appearance of $U_l$ may save the model.  We find that, unfortunately, that with real parameters in $P$ which implies $\sigma_{2,3}$ to be zero, $\delta$ and $\theta_{23}$ are predicted to be $-\pi/2$ and $\pi/4$, but there is no solutions within the allowed ranges which can predict $s_{12}$  to be consistent with data. With complex parameters in $P$, the values for $s_{12,13, 23}$ can be in agreement with data, it is, however, not possible to get $\delta$ to be close to $-\pi/2$. More numerical details will be provided in the next section.

\section{Numeral analysis}

In this section, we compare experimental data with the model predictions for the mixing angles and CP violating phase. There are several global fits of neutrino data\cite{valle,fogli-schwetz}. The latest fit gives the central values, 1$\sigma$ errors and the 2$\sigma$ ranges as the following\cite{valle}
\begin{eqnarray}
\begin{array}{ccccc}
&\delta/\pi&s_{12}^2&s_{13}^2/10^{-2}&s_{23}^2\\
NH&1.41^{+0.55}_{-0.44}&0.323\pm 0.016 &2.26\pm 0.12&0.567^{+0.032}_{-0.124}\\
2\sigma\;\mbox{region}&0.0\sim 2.0&0.292 \sim 0.357&2.02 \sim 2.50&0.414\sim 0.623\\
IH&1.48\pm 0.31&0.323\pm0.016&2.29\pm0.12&0.573^{+0.025}_{-0.039}\\
2\sigma\;\mbox{region}&0.00\sim 0.09 \& 0.86 \sim 2.0&0.292\sim 0.357&2.05\sim 2.52&0.435\sim 0.621
\end{array}
\end{eqnarray}
and the corresponding values for mass-squared differences are given by
\begin{eqnarray}
\begin{array}{ccc}
&\Delta m^2_{21} [10^{-5}] eV^2&|\Delta m^2_{31}| [10^{-3} eV^2]\\
NH&7.60^{+0.19}_{-0.18}&2.48^{+0.05}_{-0.07}\\
2\sigma\;\mbox{region}&7.26 \sim 7.99&2.35\sim 2.59\\
IH&7.60^{+0.19}_{-0.18}&2.38^{+0.05}_{-0.06}\\
2\sigma\;\mbox{region}&7.26\sim 7.99&2.26\sim 2.48
\end{array}
\end{eqnarray}
Here $NH$ and $IH$ indicate neutrino mass hierarchy patterns of normal hierarchy and inverted hierarchy, respectively. We will use the above data for comparison.

\subsection{Some general comments}

The model discussed achieved a natural way to have the CP violating angle $\delta$ to be $\pm\pi/2$ and $\theta_{23}$ to be $\pi/4$. The condition is to have all the parameters in the set $P$ to be real. This requirement can be taken as a consequence of having CP violation only caused by intrinsic source of CP violation. The condition for choosing $\delta = -\pi/2$ can be determined using eq.(\ref{j-parameter})
which requires $Im(V_{11}V_{12}^*V_{21}^*V_{22})$ to be negative.

The value $-\pi/2$  predicted in the model is in agreement with IH within 1$\sigma$ range. Although for NH case $\delta$ is outside of 1$\sigma$ range, there is no problem with 2$\sigma$ range. For $s_{23}$, the model predicts $s^2_{23} = 0.5$. This value is outside of 1$\sigma$ range for both the NH and IH cases. However, they are, again, in agreement with data within 2$\sigma$. Other parameters, $\Delta m^2_{ij}$, $s_{13}$ and $s_{12}$ can be easily fitted to within 1$\sigma$ level of current experimental data.

With complex parameters in $P$, $\delta=-\pi/2$ and $\theta_{23} = \pi/4$ can also deviate away from
$-\pi/2$ and $\pi/4$ with $\sin\delta$ determined by eq.(\ref{j-parameter}) and
$\theta_{23}$ determined by eq.(\ref{angle23}). One can find solutions where $\theta_{12, 13, 23}$ and $\delta$ to take the central values from current data.

As the general model is expected to be able to fit data, it may be more instructive to analyze some simplified versions than just providing with numbers. We provide more details of Model A and Model B discussed earlier next to see how additional assumptions restrict the level of model agreement with data.

\subsection{Model A predictions}

With real values for $w_i$ and $y$ in Model A, by adjusting the values, $w_i$, and $y$ both NH and IH mass patterns can be obtained. The predictions for $\delta$ and $\theta_{23}$ are $\pm \pi/2$ and $\theta_{23} = \pi/4$, just like the general case discussed in the previous section. Since $\delta$ should be close to $-\pi/2$, one should take the parameter space so that $c^2>s^2$.
In the model $s_{13} = (1-2cs)^{1/2}/\sqrt{3}$ is not predicted. But one can use information from $s_{13}$ to fix
$cs = 0.497\pm 0.018$  to predict $s_{12}^2= 0.334\pm0.004$ for both NH and IH cases. This is in agreement with data within 1$\sigma$. Note that $V_{e2}^2 = (s_{12}c_{13})^2 = 1/3$. It agrees with data within 1$\sigma$.

It is remarkable that neutrino mixing matrix in this model with just one free parameter can be in reasonable agreement with data. This may be a hint that it is the form for mixing matrix, at least as the lowest order approximation, that a underlying theory is producing. One should take this mass matrix seriously in theoretical model buildings.

If the parameters in the set $P$ are complex, and therefore a new phase $\rho$ appears in the model.
In this case, the new parameter $\rho$ can be used to improve agreement of the model with data.
In both NH and IH cases, $\delta$ and $s_{23}$ can be brought into agreement with data at 1$\sigma$ level.
As an example, we take the largest value of $cs$ so that $s_{13}$ takes its lower 1$\sigma$ allowed value, and then varying $\cos\rho$ to obtain the upper 1$\sigma$ allowed value. This fixes $cs$ and $\cos\rho$ to be
0.468 and 0.992, respectively. With these values, $s_{23}$ and $\delta$ are determined to:
$0.534$ and $1.426 \pi$, respectively. These values are in agreement with data at 1$\sigma$ level.

\begin{figure}[!h]
\begin{centering}
\begin{tabular}{cc}
\includegraphics[width=0.3\textwidth]{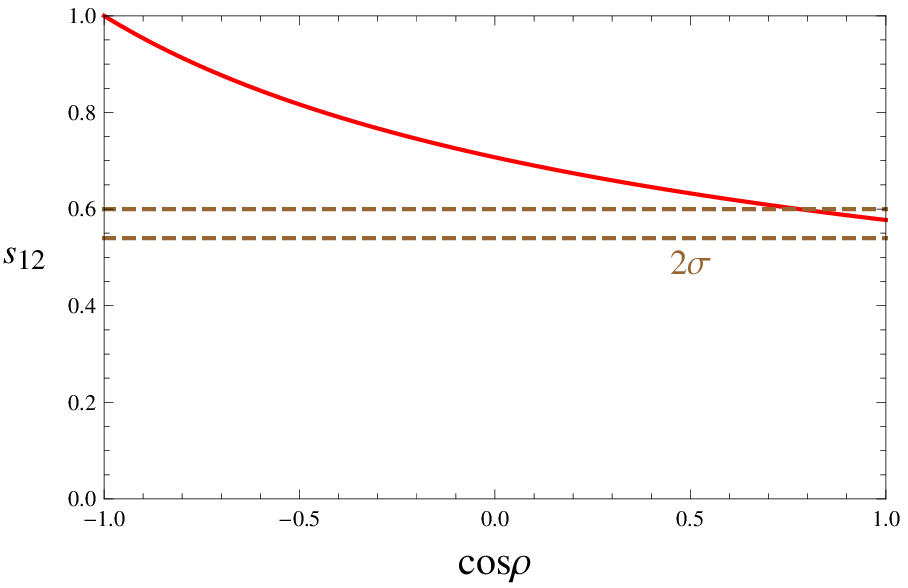}
\hspace*{1em}
\includegraphics[width=0.3\textwidth]{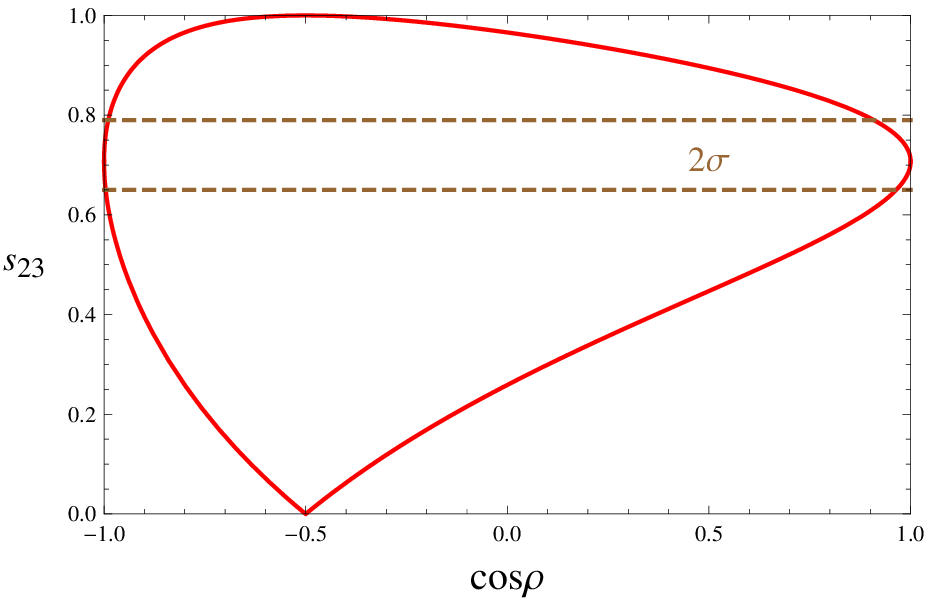}
\hspace*{1em}
\includegraphics[width=0.3\textwidth]{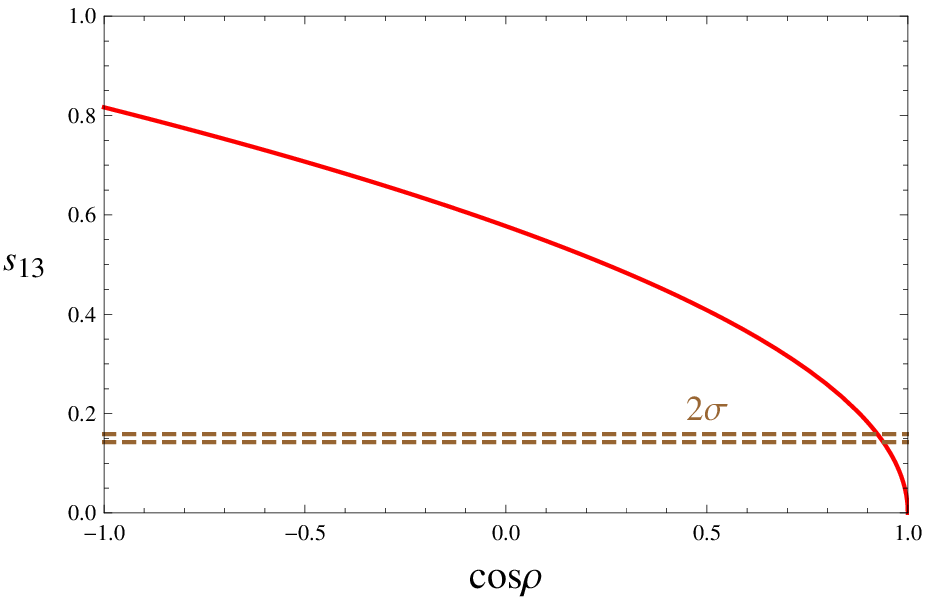}
\end{tabular}
\end{centering}
\caption{$s_{12}$, $s_{23}$, $s_{13}$ as  functions of $cos\rho$ with $c=s=\sqrt{2}/2$  compared with allowed ranges for $(s_{13},\;s_{23},\;s_{12})$ at 1$\sigma$ and 2$\sigma$, respectively, given by
$(0.147 \sim 0.155,\;0.731 \sim 0.773,\;0.554 \sim 0.582)$ and $(0.143 \sim 0.159,\;0.660 \sim 0.788,\;0.540 \sim 0.597)$.}\label{modelA}
\end{figure}

For the case with $c=s$, the model is more restrictive. In this case $\sin\delta =0$. With more precise data on the CP violating angle $\delta$, this may rule out this simple case with high confidence level. If $\rho=0$, the model is already ruled out at high precision from $s_{13}$ measurement. However, with a non-zero $\rho$, the mixing angles can still be made to in agreement at 2$\sigma$ level. 
In Fig. \ref{modelA} we show
$s_{12}$, $s_{23}$, $s_{13}$ as functions of $\rho$. When chose $cos\rho=0.93$, we can get $s_{12}=0.58, s_{23}=0.78, s_{13}=0.15$ which agree with the experimental data within $2\sigma$ range.

More precise experimental data are required to distinguish the model with complex model parameters from that with the real parameters and other models, or to rule out the above simples completely.

\subsection{Model B predictions}

\begin{figure}[h!]
\begin{centering}
\begin{tabular}{c}
\includegraphics[width=6.5cm]{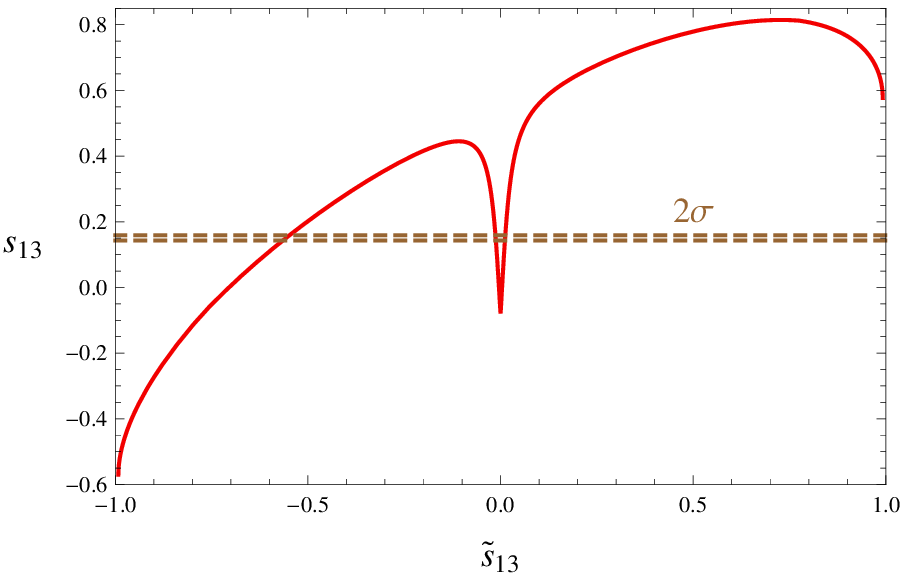}
\hspace*{2em}
\includegraphics[width=6.5cm]{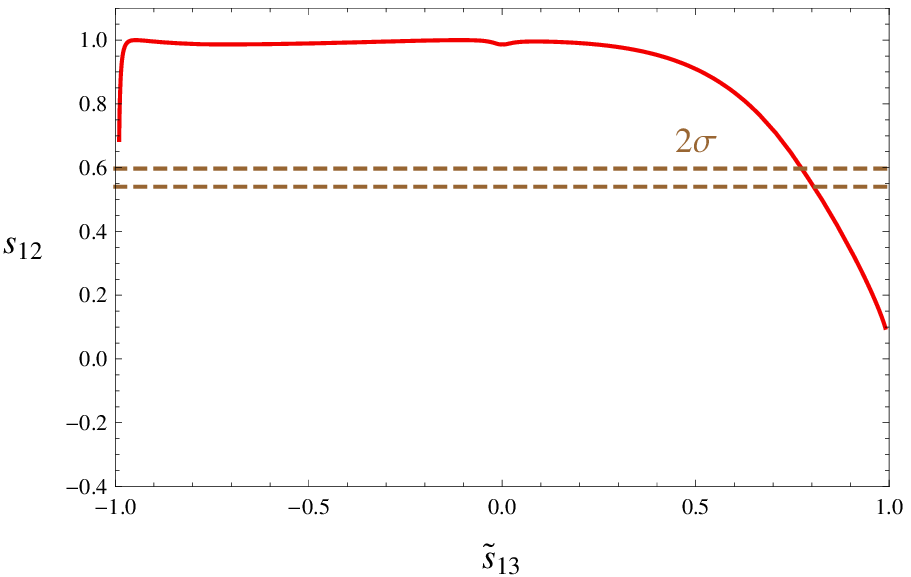}
\end{tabular}
\caption{$s_{13}$, $s_{12}$ as functions of  $\tilde s_{13}$.  }\label{modelB1}
\end{centering}
\end{figure}

With real parameters in $P$ in this model, the predictions, $\delta$ and $\theta_{23}$ are the same as the general case. Using eq.(\ref{model-b}), one can choose $\tilde s_{13}$ to be the parameter to fit all data. We find that no solutions can simultaneously bring $s_{13}$ and $s_{12}$ to be in 2$\sigma$ allowed region compared with data. In Fig. \ref{modelB1} we show $s_{13}$, $s_{12}$ as functions of $\tilde s_{13}$.
Here, when $s_{13}$ agree with the data in $2\sigma$ range, the allowed region for $\tilde s_{13}$ are
$(-0.594 \sim -0.565) \& (-0.012 \sim -0.010)\& (0.010 \sim 0.011)$, but when $s_{12}$ agrees with the data in the $2\sigma$ ranges,
the allowed region for $\tilde s_{13}$ are only $(0.806 \sim 0.774)$, they have no overlap region for $\tilde s_{13}$. This model is therefore ruled out.

\begin{figure}[h!]
 \centering
\begin{minipage}[t]{6.24cm}
\centering
\includegraphics[width=6.5cm]{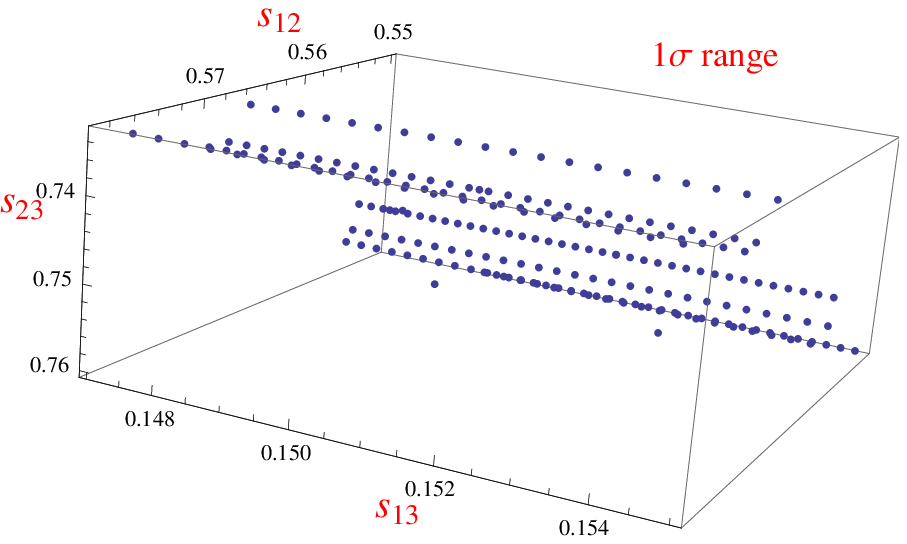}
\end{minipage}
\hspace*{2em}
\begin{minipage}[t]{6.24cm}
\centering
\includegraphics[width=6.5cm]{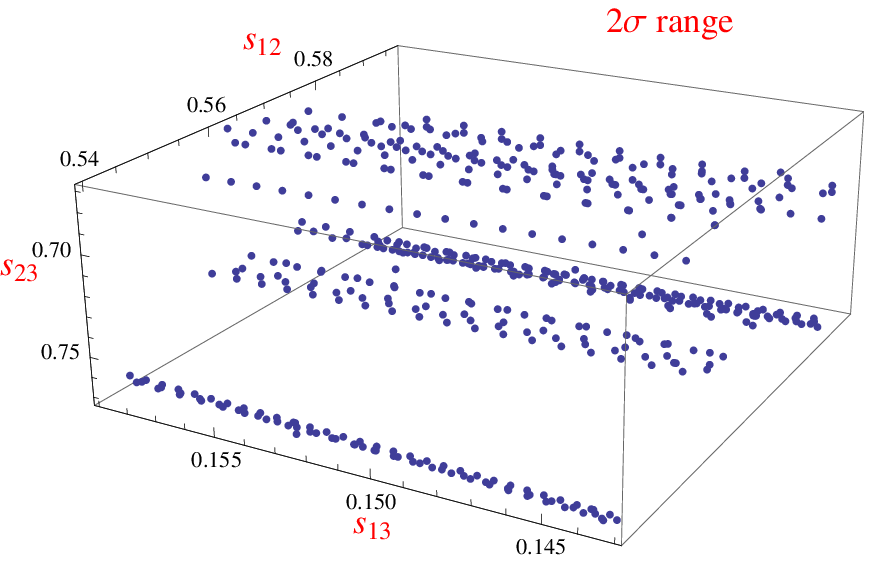}
\end{minipage}
\caption{The scanning points of the three free parameters in $1\sigma$ and $2\sigma$ region respectively.  }\label{modelB2}
\end{figure}

With complex parameters in $P$,  there are solutions of $\tilde s_{13}$ and $\sigma_i$ so that to make $s_{23}$, $s_{13}$ and $s_{12}$ to be consistent with data at 1$\sigma$ level. However this requires one of  $|\sin\sigma_i|$ to be away from zero which alter the prediction for $\delta$ significantly away from $-\pi/2$, although $\theta_{23}$ is still close to $\pi/4$.
In Fig. \ref{modelB2}, we scan the parameter space $-1\leq \tilde s_{13}\leq 1,  0\leq\sigma_2\leq2\pi,  0\leq\sigma_3\leq2\pi$, and obtained
the corresponding values of $s_{12}$, $s_{23}$, $s_{13}$ that agree with data within  $1\sigma$ and $2\sigma$ region respectively. And in $1\sigma$ range, the CP phase will be constrained in $sin\delta\sim(-0.5,0.5)$, but in $2\sigma$ region, CP phase is free and can be in the range of  $0$ to $2\pi$.

For example with
$\tilde s_{13}=0.57$, $\sigma_{2}={7\pi/5}$, $\sigma_{3}={26\pi / 25}$, one can obtain that
$s_{13}=0.15$, $s_{23}=0.74$, $s_{12}=0.57$. These mixing angles are within the 1$\sigma$ allowed region, but it will have a  CP phase $\delta=1.06 \pi$ significantly away from $3\pi/2$. A precise determination of the CP violating phase is needed to rule out the above model.

\section{Conclusions}

We have constructed theoretical models for neutrino mass matrix in Type-II seesaw with $A_4$ family symmetry.
The models we constructed naturally predict that the CP violating phase $\delta$ is equal to $\pm\pi/2$ and at the same time the mixing angle $\theta_{23}$ is $\pi/4$.   The reality of the parameters can be achieved by imposing a generalized CP symmetry in the Grimus-Lavoura symmetry limit.
When the generalized CP symmetry is explicitly broken, the Yukawa couplings can be complex, the model predictions for $\delta$ and $\theta_{23}$ can be significantly modified. Two simple scenarios, Model A and Model B, are analyzed in detail to show how the modifications arise and how future experimental data can test this class of models.

The Model A has a characteristic prediction that $|V_{e2}|=1/\sqrt{3}$. This model can accommodate experimental data at 1$\sigma$ level. It can be taken as the lowest order neutrino mass matrix for future theoretical model buildings. When Yukawa couplings are complex,  the CP violating phase $\delta$ and the mixing angle $\theta_{23}$ can be away from $-\pi/2$ and $\pi/4$. The crucial test for this model is to measure whether $|V_{e2}|=1/\sqrt{3}$ holds to high precision.

In Model B, the diagonal entries of neutrino mass matrix are all zero in the weak interaction basis. This implies that  the neutrino masses can be determined by the known neutrino mass-squared differences. We find that this model can only accommodate inverted neutrino mass hierarchy. The mixing angles $s_{12}$, $s_{13}$ and $s_{23}$ cannot simultaneously be in agreement with data at 2$\sigma$ level if the Yukawa couplings are all real. With complex Yukawa couplings, the mixing angles can be brought into agreement with data at 1$\sigma$ level, but the CP violating angle $\delta$ will be significantly away from $-\pi/2$. This provides a crucial test for this model.

At present, experimental data on neutrino mixing seem to hint that  $\delta$ and $\theta_{23}$ to be close to $-\pi/2$ and $\pi/4$. Theoretical models which can naturally achieve such predictions are interesting to study. The models we have constructed have many novel properties and can be tested. We will have to wait future  experimental data to tell us more whether this class of neutrino mass matrix will survive.

\begin{acknowledgments}

The work was supported in part by MOE Academic Excellent Program (Grant No: 102R891505) and MOST of ROC, and in part by NSFC(Grant No:11175115) and Shanghai Science and Technology Commission (Grant No: 11DZ2260700) of PRC.

\end{acknowledgments}

\end{document}